\documentclass[aps,pra,10pt,showpacs,oneside,preprint,twocolumn,letterpaper,superbib,notitlepage]{revtex4}
\usepackage{graphicx}
\usepackage{amsfonts}
\usepackage{amsmath}

\input{tcilatex}

\begin{document}

\title{Quadrature and polarization squeezing in a dispersive optical
bistability model}
\author{Ferran V. Garcia--Ferrer$^{1}$, Isabel P\'{e}rez--Arjona$^{2}$, Germ%
\'{a}n J. de Valc\'{a}rcel$^{1}$, and Eugenio Rold\'{a}n$^{1}$}
\affiliation{$^{1}$Departament d'\`{O}ptica, Universitat de Val\`{e}ncia, Dr. Moliner 50,
46100--Burjassot, Spain}
\affiliation{$^{2}$Departament de F\'{\i}sica Aplicada, Escola Polit\`{e}cnica Superior
de Gandia, Universitat Polit\`{e}cnica de Val\`{e}ncia, Ctra. Nazaret--Oliva
S/N, 46730--Grau de Gandia, Spain}

\begin{abstract}
We theoretically study quadrature and polarization squeezing in dispersive
optical bistability through a vectorial Kerr cavity model describing a
nonlinear cavity filled with an isotropic $\chi ^{\left( 3\right) }$ medium
in which self-phase and cross-phase modulation, as well as four--wave
mixing, occur. We derive expressions for the quantum fluctuations of the
output field quadratures as a function of which we express the spectrum of
fluctuations of the output field Stokes parameters. We pay particular
attention to study how the bifurcations affecting the non-null linearly
polarized output mode squeezes the orthogonally polarized vacuum mode, and
show how this produces polarization squeezing.
\end{abstract}

\maketitle

\section{Introduction}

Squeezing has received continued attention since the early eighties \cite%
{Loudon87,Meystre91,Drummond04} because it is a beautiful quantum phenomenon
with amazing potential applications \cite{Drummond04}, the most recent of
which are connected with continuous variable quantum information \cite%
{Braunstein05}. One particular aspect of squeezing that is receiving more
attention at present is polarization squeezing, probably because its
detection does not need homodyning, as it is the case with quadrature
squeezing, i.e., photodetectors suffice for detecting polarization squeezing 
\cite{Korolkova02,Heersink03,Luis06}. Here we are concerned with the study
of both quadrature and polarization squeezing in the field exiting a
particular nonlinear optical cavity.

One system that attracted early attention for quadrature squeezing is
dispersive optical bistability \cite%
{Drummond80b,Vogel,LugiatoOB,Orozco87,WallsMilburn} which consists in an
optical cavity filled with a nonlinear $\chi ^{\left( 3\right) }$ medium
which is fed with an external (pumping) field. For appropriate values of the
system parameters, the output field exhibits bistability and at the turning
points of the hysteresis cycle, perfect squeezing is obtained. Here we
revisit this system because we are interested not only in quadrature but
also in polarization squeezing.

The generalization of this simple dispersive optical bistability model we
consider in this article consists in taking into account the existence,
inside the optical cavity, of a mode with orthogonal polarization with
respect to that of the pumping field, which is also close to cavity
resonance. For low pump, this extra mode remains empty, i.e., the output
cavity field has the same polarization as the pumping field, but for large
enough pump this orthogonally polarized mode can switch on (what is known as
polarization instability) and then the field exiting the cavity changes from
linearly to elliptically polarized. This system is known as vectorial Kerr
cavity \cite{Geddes94} and its mathematical modeling consists in taking into
account self-- and cross--phase modulation, as well as four--wave mixing,
for the two orthogonally polarized intracavity modes, the weight of each of
the three nonlinear processes depending on the particular nonlinear medium
under consideration \cite{Boyd}.

The vectorial Kerr cavity has been studied in the past from the viewpoints
of stability theory \cite{Sanchez-Morcillo00} and pattern formation \cite%
{Geddes94,Hoyuelos98,Gallego00,Sanchez00b,Perez00}. Also quantum
fluctuations have been studied in this system but, as far as we know, only
from the perspective of pattern formation \cite{Zambrini02}, i.e., by
considering a continuum of transverse modes. Here we consider the single
transverse mode model and study quadrature and polarization squeezing in the
field exiting the nonlinear cavity. As we show below, the vectorial Kerr
cavity exhibits polarization squeezing. In particular, and most
interestingly, this polarization squeezing is optimum at the bistability
bifurcation points, when the orthogonally polarized field is off. Thus the
vectorial Kerr cavity model, which can be taken as a more accurate model of
dispersive optical bistability, reveals that this well known process can act
as a polarization squeezer.

The rest of the article is structured as follows. In Section II\ we derive
the Langevin equations for quantum fluctuations from the quantum Hamiltonian
of a Kerr cavity pumped by a linearly polarized coherent field, which are
derived from the Fokker--Planck equation verified by the generalized-P
distribution. In Sec. III, in order to keep in mind the scenario of the
different solutions exhibited by the system, their classical stationary
solutions are given and their stability properties are briefly reviewed. The
linearized Langevin equations around the semiclassical stationary solutions
and the spectrum of quantum fluctuations are introduced in Sec. IV. Sec. V
is devoted to the analysis of quadrature squeezing. Then, in Section VI
quantum Stokes operators and the definition of polarization squeezing are
introduced. In Sec. VII we state specifically the relation between the
Stokes parameters variances and the quadrature squeezing spectra and analyze
polarization squeezing in this system, paying special attention to the
squeezing occurring at the bifurcations affecting the linearly polarized
solution, but considering also polarization squeezing in the elliptically
polarized solution in a particular case. Finally, in Sec. VIII we show the
most relevant conclusions of our work.

\section{Model}

Consider an optical cavity filled with an isotropic Kerr medium and assume
that cavity losses occur only at one of the cavity mirrors being the same
for all possible field polarizations (i.e., assume that the cavity is also
isotropic). An external laser field, whose frequency is far enough from any
resonance in the nonlinear medium (what makes possible that the interaction
be described by a real cubic nonlinear susceptibility) is injected into the
cavity. We shall assume that the injected field is linearly polarized, say
along the $x$-direction. As cavity losses do not favour the injected field
polarization, under certain circumstances \cite{Sanchez-Morcillo00} the
intracavity field can experience a polarization instability leading to the
spontaneous generation of a field orthogonally polarized with respect to the
injected field ($y$-polarized in our case), hence the name vectorial Kerr
cavity for such a nonlinear system.

The quantum Hamiltonian which describes this system is given, in the
interaction picture, by 
\begin{equation}
\hat{H}=\hat{H}_{free}+\hat{H}_{ext}+\hat{H}_{int},  \label{Hamiltonian}
\end{equation}%
with 
\begin{subequations}
\begin{eqnarray}
\hat{H}_{free} &=&\hbar \left( \omega _{c}-\omega _{0}\right) \left( \hat{a}%
_{1}^{\dagger }\hat{a}_{1}+\hat{a}_{2}^{\dagger }\hat{a}_{2}\right) , \\
\hat{H}_{ext} &=&i\hbar E_{0}\left( \hat{a}_{1}^{\dagger }-\hat{a}%
_{1}\right) , \\
\hat{H}_{int} &=&-\hbar \eta g\frac{1}{2}\left( \hat{a}_{1}^{\dagger 2}\hat{a%
}_{1}^{2}+\hat{a}_{2}^{\dagger 2}\hat{a}_{2}^{2}\right) \\
&&-\hbar \eta g\left[ \mathcal{A}\hat{a}_{1}^{\dagger }\hat{a}_{1}\hat{a}%
_{2}^{\dagger }\hat{a}_{2}+\frac{\mathcal{B}}{4}\left( \hat{a}_{1}^{2}\hat{a}%
_{2}^{\dagger 2}+\hat{a}_{1}^{\dagger 2}\hat{a}_{2}^{2}\right) \right] , 
\notag
\end{eqnarray}%
where $\hat{H}_{free}$ corresponds to the free Hamiltonian of the two
orthogonally polarized intracavity modes, $\hat{H}_{ext}$ describes the
injected coherent field, and $\hat{H}_{int}$ is the interaction Hamiltonian.
In the above expressions $\hat{a}_{i}^{\dagger }$ and $\hat{a}_{i}$ are,
respectively, the creation and annihilation operators corresponding to the $%
x $-polarized ($i=1$) and the $y$-polarized ($i=2$) fields; $\omega _{0}$
and $\omega _{c}$ are the injected field frequency and the frequency of the
cavity mode closest to $\omega _{0}$ (which is assumed to be the same for
the two orthogonally polarized modes as the cavity is isotropic),
respectively; $E_{0}$ is proportional to the amplitude of the injected
coherent field (which we take as a real quantity without loss of
generality); $\mathcal{A}$ and $\mathcal{B}$ are the Maker--Terhune
coefficients (which, for isotropic media, satisfy $\mathcal{A}+\mathcal{B}%
/2=1$ \cite{Boyd}) governing the relative strength of cross--phase
modulation and four--wave mixing, respectively; $g=3\varepsilon _{0}\hbar
\omega _{c}^{2}\chi /\varepsilon ^{2}V$ is the radiation--matter coupling
constant ($V$ is the cavity volume, $\varepsilon $ is the medium dielectric
constant and $\chi =\chi _{iiii}^{(3)}$, $i=x,y,z$,\ is the nonlinear
susceptibility \cite{Boyd}); and, finally, $\eta =\pm 1$ takes account of
the self-focusing ($\eta =+1$) or self-defocusing ($\eta =-1$) cases.

The intracavity field exits the cavity through the single output mirror.
Treating, as it is usual, the external vacuum modes as a reservoir \cite%
{Carmichael}, the master equation governing the evolution of density matrix $%
\rho $ of the intracavity modes is 
\end{subequations}
\begin{subequations}
\begin{eqnarray}
\frac{\partial }{\partial t}\hat{\rho} &=&\frac{1}{i\hbar }[\hat{H},\hat{\rho%
}]+\widehat{\Lambda \rho },  \label{Master Eq} \\
\widehat{\Lambda \rho } &=&\gamma \sum_{i=1,2}\left( [\hat{a}_{i},\hat{\rho}%
\hat{a}_{i}^{\dagger }]+[\hat{a}_{i}\hat{\rho},\hat{a}_{i}^{\dagger
}]\right) ,  \label{liouvillian}
\end{eqnarray}%
the Liouvillian term $\widehat{\Lambda \rho }$ modeling the coupling between
the system and the external reservoir through the output mirror ($\gamma $
denotes cavity losses; $2\gamma $ is the photon number loss rate).

\subsection{Fokker--Planck and Langevin equations}

In order to obtain Langevin equations describing c-numbers evolution from
the master equation (\ref{Master Eq}) one needs an appropriate quantum
quasiprobability distribution. We choose the generalized P-representation 
\cite{Drummond80}, which is the most convenient in our case as the output
field correlations are most easily derived from the intracavity ones with
this representation: On one hand, it is suitable for obtaining correlations
of normally ordered products of operators -which provides the calculation of
fluctuations spectra outside the resonator- while, on the other hand, avoids
the problems associated with other quasiprobablity distributions, such as
the Glauber-P representation, which although also gives normally ordered
correlation products, can exhibit non positive--semidefinite diffusion
matrix in its dynamical equation.

The generalized-P representation sets a correspondence between quantum
operators $\hat{a}_{i}(t)$ and $\hat{a}_{i}^{\dagger }(t)$ and independent
c-numbers $\alpha _{i}(t)$ and $\alpha _{i}^{+}(t)$, respectively,
satisfying in their mean value that $\left\langle \alpha
_{i}(t)\right\rangle ^{\ast }=\left\langle \alpha _{i}^{+}(t)\right\rangle $%
. We have derived the equation of evolution for the generalized-P
distribution $P\left( \mathbf{a},t\right) $ by using standard techniques 
\cite{WallsMilburn}, and have obtained the following well behaved
Fokker--Planck equation 
\end{subequations}
\begin{equation}
\frac{\partial P}{\partial t}=\left[ -\underset{i=1}{\overset{4}{\sum }}%
\frac{\partial }{\partial \alpha _{i}}A_{i}\left( \mathbf{a}\right) +\tfrac{1%
}{2}\underset{i,j=1}{\overset{4}{\sum }}\frac{\partial ^{2}}{\partial \alpha
_{i}\partial \alpha _{j}}D_{i,j}\left( \mathbf{a}\right) \right] P,
\label{FokkerPlanck}
\end{equation}%
with $\mathbf{a}=\func{col}(\alpha _{1},\alpha _{1}^{+},\alpha _{2},\alpha
_{2}^{+})$ and 
\begin{subequations}
\begin{eqnarray}
A_{1}(\mathbf{a}) &=&E_{0}-\gamma \left( 1+i\eta \Delta \right) \alpha
_{1}+i\eta g\times \\
&&\times \left[ \alpha _{1}^{2}\alpha _{1}^{+}+\mathcal{A}\alpha _{1}\alpha
_{2}\alpha _{2}^{+}+\frac{\mathcal{B}}{2}\alpha _{1}^{+}\alpha _{2}^{2}%
\right] ,  \notag \\
A_{2}(\mathbf{a}) &=&E_{0}-\gamma \left( 1-i\eta \Delta \right) \alpha
_{1}^{+}-i\eta g\times \\
&&\times \left[ \alpha _{1}\left( \alpha _{1}^{+}\right) ^{2}+\mathcal{A}%
\alpha _{1}^{+}\alpha _{2}\alpha _{2}^{+}+\frac{\mathcal{B}}{2}\alpha
_{1}\left( \alpha _{2}^{+}\right) ^{2}\right] ,  \notag \\
A_{3}(\mathbf{a}) &=&-\gamma \left( 1+i\eta \Delta \right) \alpha _{2}+i\eta
g\times \\
&&\times \left[ \alpha _{2}^{2}\alpha _{2}^{+}+\mathcal{A}\alpha _{1}\alpha
_{2}\alpha _{1}^{+}+\frac{\mathcal{B}}{2}\alpha _{1}^{2}\alpha _{2}^{+}%
\right] ,  \notag
\end{eqnarray}%
\begin{eqnarray}
A_{4}(\mathbf{a}) &=&-\gamma \left( 1-i\eta \Delta \right) \alpha
_{2}^{+}-i\eta g\times \\
&&\times \left[ \alpha _{2}\left( \alpha _{2}^{+}\right) ^{2}+\mathcal{A}%
\alpha _{1}\alpha _{2}^{+}\alpha _{1}^{+}+\frac{\mathcal{B}}{2}\left( \alpha
_{1}^{+}\right) ^{2}\alpha _{2}\right] ,  \notag
\end{eqnarray}%
where the new detuning parameter $\Delta $ is

\end{subequations}
\begin{equation}
\Delta =\frac{1}{\eta \gamma }\left[ \omega _{c}-\omega _{0}-\eta g\left( 1+%
\frac{\mathcal{A}}{2}\right) \right] =\frac{1}{\eta \gamma }\left( \bar{%
\omega}_{c}-\omega _{0}\right) .
\end{equation}
with $\bar{\omega}_{c}=\omega _{c}-\eta g\left( 1+\mathcal{A}/2\right) $ the
shifted cavity mode frequency (the shift appears because of the vacuum
fluctuations contribution, which depends on the ordering chosen when writing
the Hamiltonian \cite{Garcia05}). Finally, the diffusion matrix elements are 
\begin{subequations}
\label{DiffusionMatrix}
\begin{eqnarray}
D_{11} &=&i\eta g\left( \alpha _{1}^{2}+\frac{\mathcal{B}}{2}\alpha
_{2}^{2}\right) , \\
\ D_{22} &=&-i\eta g\left[ \left( \alpha _{1}^{+}\right) ^{2}+\frac{\mathcal{%
B}}{2}\left( \alpha _{2}^{+}\right) ^{2}\right] , \\
D_{33} &=&i\eta g\left( \alpha _{2}^{2}+\frac{\mathcal{B}}{2}\alpha
_{1}^{2}\right) , \\
D_{44} &=&-i\eta g\left[ \left( \alpha _{2}^{+}\right) ^{2}+\frac{\mathcal{B}%
}{2}\left( \alpha _{1}^{+}\right) ^{2}\right] , \\
D_{13} &=&D_{31}=-i\eta g\mathcal{A}\alpha _{1}\alpha _{2}, \\
D_{24} &=&D_{42}=i\eta g\mathcal{A}\alpha _{1}^{+}\alpha _{2}^{+},
\end{eqnarray}%
being null the rest of elements.

The Fokker-Planck equation (\ref{FokkerPlanck}) can be transformed into an
equivalent set of classical--looking stochastic differential Langevin
equation via the Ito rules \cite{Gardiner00}. These Langevin equations read 
\end{subequations}
\begin{equation}
\frac{d\mathbf{a}}{dt}=\mathbf{A}\left( \mathbf{a}\right) +B\left( \mathbf{a}%
\right) \cdot \boldsymbol{\xi }\left( t\right) ,  \label{LangevinCompleta}
\end{equation}%
where $\mathbf{A}=\func{col}\left( A_{1},A_{2},A_{3},A_{4}\right) $, $%
\boldsymbol{\xi }=\func{col}\left( \xi _{1},\xi _{1}^{+},\xi _{2},\xi
_{2}^{+}\right) $, with $\xi _{i}(t)$ and $\xi _{i}^{+}(t)$ independent
white Gaussian noises of zero average and correlations verifying 
\begin{subequations}
\label{NoiseProperties}
\begin{eqnarray}
\left\langle \xi _{i}(t),\xi _{j}^{+}(t^{\prime })\right\rangle &=&0, \\
\left\langle \xi _{i}^{+}(t),\xi _{j}^{+}(t^{\prime })\right\rangle
&=&\delta _{ij}\delta (t-t^{\prime }), \\
\left\langle \xi _{i}(t),\xi _{j}(t^{\prime })\right\rangle &=&\delta
_{ij}\delta (t-t^{\prime }),
\end{eqnarray}%
and matrix $B\left( \mathbf{a}\right) $ satisfying

\end{subequations}
\begin{equation}
D(\mathbf{a})=B\left( \mathbf{a}\right) \cdot B^{T}\left( \mathbf{a}\right) .
\label{DefB}
\end{equation}

\section{Classical model}

Let us write the model equations (\ref{LangevinCompleta}) in the classical
limit (i.e., when $\alpha _{i}^{+}$ is interpreted as $\alpha _{i}^{\ast }$
and the noise terms are ignored). They read 
\begin{subequations}
\label{clas}
\begin{eqnarray}
\frac{d\alpha _{1}}{dt} &=&E_{0}-\gamma \left( 1+i\eta \Delta \right) \alpha
_{1}  \label{clas1} \\
&&+i\eta g\left[ \left\vert \alpha _{1}\right\vert ^{2}\alpha _{1}+\mathcal{A%
}\left\vert \alpha _{2}\right\vert ^{2}\alpha _{1}+\frac{\mathcal{B}}{2}%
\alpha _{1}^{\ast }\alpha _{2}^{2}\right] ,  \notag \\
\frac{d\alpha _{2}}{dt} &=&-\gamma \left( 1+i\eta \Delta \right) \alpha _{2}
\label{clas2} \\
&&+i\eta g\left[ \left\vert \alpha _{2}\right\vert ^{2}\alpha _{2}+\mathcal{A%
}\left\vert \alpha _{1}\right\vert ^{2}\alpha _{2}+\frac{1}{2}\mathcal{B}%
\alpha _{1}^{2}\alpha _{2}^{\ast }\right] .  \notag
\end{eqnarray}%
The above equations exhibit the symmetry $\left\{ \alpha _{1},\alpha
_{2},\Delta ,\eta \right\} \leftrightarrow \left\{ \alpha _{1}^{\ast
},\alpha _{2}^{\ast },-\Delta ,-\eta \right\} $ (remember that $E_{0}$ has
been taken to be real without loss of generality). Then passing from
positive $\eta $ (self--focusing)\ to negative $\eta $ (self--defocusing)\
is equivalent to changing $\Delta \leftrightarrow -\Delta $ for what
concerns the steady states and their stability properties.

By introducing the following changes 
\end{subequations}
\begin{subequations}
\label{ReNormalizations}
\begin{equation}
\tau =\gamma t,\ \ \ E=\frac{1}{\gamma }\sqrt{\frac{g}{\gamma }}E_{0},\ \ \
a_{i}\equiv \sqrt{\frac{g}{\gamma }}\alpha _{i},  \label{a}
\end{equation}%
Eqs. (\ref{clas}) transform into an equivalent set of equations coinciding
with those given in \cite{Sanchez-Morcillo00}. Then we can use the results
previously derived in \cite{Sanchez-Morcillo00} by taking into account the
above changes. (We notice that along the following sections it will be more
convenient to give some of the expressions in terms of $a_{i}$, $E$, and $%
\tau $ than in terms of the original quantities, so we shall refer later to
the above transformations).

Eqs. (\ref{clas}) have two different steady state solutions (see \cite%
{Sanchez-Morcillo00} and references therein): The singlemode (or linearly
polarized) solution, which corresponds to the pure Kerr solution \cite%
{LugiatoOB} 
\end{subequations}
\begin{subequations}
\label{monomode}
\begin{eqnarray}
E_{0}^{2} &=&\gamma ^{2}\left[ 1+\left( \Delta -\frac{g}{\gamma }%
I_{1s}\right) ^{2}\right] I_{1s},  \label{monomode-sol} \\
\phi _{1s} &=&\arccos \left( \gamma \frac{\sqrt{I_{1s}}}{E_{0}}\right)
\label{monomode-sol2} \\
I_{2s} &=&0,  \label{monomode-sol3}
\end{eqnarray}%
where $\alpha _{j}=\sqrt{I_{js}}e^{i\phi _{js}}$ at steady state ($j=1,2$);
and the bimode (or elliptically polarized) solution 
\end{subequations}
\begin{subequations}
\label{Bimode}
\begin{eqnarray}
E_{0}^{2}I_{1s} &=&\gamma ^{2}(I_{1s}+I_{2s})^{2}  \label{Bimode-sol1} \\
&&+g^{2}(I_{1s}-I_{2s})^{2}(I_{1s}+I_{2s}-\frac{\gamma \Delta }{g})^{2}, 
\notag \\
I_{2s} &=&\frac{\gamma \Delta }{g}-\mathcal{A}I_{1s}\pm \sqrt{\left( \frac{%
\mathcal{B}}{2}\right) ^{2}I_{1s}^{2}-\left( \frac{\gamma }{g}\right) ^{2}},
\label{Bimode-sol2} \\
\phi _{1s} &=&\arccos \left[ \frac{\gamma \sqrt{I_{1s}}}{E_{0}}\left( 1+%
\frac{I_{2s}}{I_{1s}}\right) \right] ,\ \ \   \label{Bimode-sol3} \\
\phi _{2s} &=&\phi _{1s}+\frac{1}{2}\arccos \left( \frac{\gamma \Delta
-gI_{1s}-\mathcal{A}gI_{2s}}{\frac{\mathcal{B}}{2}gI_{1s}}\right) .
\label{Bimode-sol4}
\end{eqnarray}%
Notice that the bimode solution with the same intensity and phase $\phi
_{2}=\phi _{2s}+\pi $ also exists, i.e., Eqs. (\ref{clas}) exhibit phase
bistability for the cross--polarized mode.

The stability of these two steady state solutions has been extensively
studied in \cite{Sanchez-Morcillo00} and turns out to be quite involved.
Fortunately for our present purposes it will suffice to consider the
stability of the singlemode solution and consider parameter values where the
bimode solution does not experience secondary bifurcations, an information
we can extract from \cite{Sanchez-Morcillo00}. From now on we shall restrict
ourselves to the case of liquids, for which $\mathcal{A}=1/4$ and $\mathcal{B%
}=3/2$, and we refer the interested reader to \cite{Sanchez-Morcillo00} for
full details and general expressions (including possible anisotropies in the
cavity losses). It is interesting to say here that there is not a threshold
value for the Maker-Terhune coefficients for the polarization instability
exists, what occurs is that the pump needed for reaching that instability
tends to infinity as $\mathcal{B}\rightarrow 0$ (a limit that corresponds to
homogeneous media in which electrostriction is the physical origin of the
nonlinearity \cite{Boyd}).

The singlemode (linearly polarized) solution $\left( \alpha _{1}\neq
0,\alpha _{2}=0\right) $, which exists for any parameter set, can undergo
two different bifurcations:\ A tangent bifurcation (\textit{bistability\
bifurcation}) associated with the appearance of optical bistability, and a
pitchfork bifurcation (\textit{polarization bifurcation}) that leads to the
switching--on of the orthogonal polarized field $\alpha _{2}$, i.e., to the
bimode solution $\left( \alpha _{1}\neq 0,\alpha _{2}\neq 0\right) $. There
are no Hopf bifurcations affecting the singlemode solution.

The bistability of the linearly polarized solution requires that $\Delta >%
\sqrt{3}$ and the multisolution region exists for pump values between $%
E_{0,-}^{2}$ and $E_{0,+}^{2}$ with 
\end{subequations}
\begin{equation}
E_{0,\pm }^{2}=\frac{2\gamma ^{3}}{27g}\left[ \Delta \left( \Delta
^{2}+9\right) \pm \eta \sqrt{\left( \Delta ^{2}-3\right) ^{3}}\right] ,
\label{bistabilityLoop}
\end{equation}%
or, equivalently, for intracavity intensity values between $I_{1s,-}$ and $%
I_{1s,+}$ with%
\begin{equation}
I_{1s,\pm }=\gamma \frac{2\Delta \pm \sqrt{\Delta ^{2}-3}}{3g}.
\label{ibistability}
\end{equation}

The polarization bifurcation, corresponding to the switching--on of the $%
\alpha _{2}$ field occurs at the pump intensity value%
\begin{equation}
E_{0,pol}^{2}=\frac{3\gamma ^{3}}{g}\left[ (\Delta ^{2}+\frac{1}{2})\left( 
\sqrt{9\Delta ^{2}+8}-3\Delta \right) -\Delta \right] ,  \label{BimodeSwitch}
\end{equation}%
or, equivalently, at the intracavity intensity value%
\begin{equation}
I_{1s,pol}=\gamma \frac{-\Delta +\sqrt{9\Delta ^{2}+8}}{2g}.
\label{ipolarization}
\end{equation}%
For pump values larger than $E_{0,pol}^{2}$ (or $I_{1s}>I_{1s,pol}$) the
solution is always the elliptically polarized one (this is true for liquids
but not a general result \cite{Sanchez-Morcillo00}) and exists for any value
of $\Delta $. In Fig. 1 the two tangent bifurcations and the polarization
instability are represented on the $\left\langle \Delta ,E^{2}\right\rangle $
plane, see Eqs.(\ref{ReNormalizations}).

As stated, the stability properties of the elliptically polarized solution
are much more involved than those of the singlemode solution: It can undergo
secondary tangent bifurcations as well as Hopf bifurcations and no
analytical expressions exist for the parameter values at these secondary
bifurcations. However these scenarios are out of the scope of the present
work as we will consider parameter values for which these secondary
bifurcations do not occur.

\section{Linearized quadrature squeezing spectra}

We return now to the quantum description of the system. In order to describe
in a simple way the dynamics of quantum fluctuations, the Langevin equations
(\ref{LangevinCompleta}) are linearized around the stationary classical mean
values of the intracavity fields, given by Eqs. (\ref{monomode}) for the
singlemode solution (only the field parallel to the injected one is present
in the cavity) and Eqs. (\ref{BimodeSwitch}) for the bimode solution (the
orthogonal polarized fields switches on and the output field is elliptically
polarized). We thus write 
\begin{subequations}
\begin{eqnarray}
\alpha _{j} &=&\sqrt{I_{js}}e^{+i\phi _{js}}+\delta \alpha _{j}, \\
\alpha _{j}^{+} &=&\sqrt{I_{js}}e^{-i\phi _{js}}+\delta \alpha _{j}^{+},
\end{eqnarray}%
with $j=1,2$ and where the $I_{js}$ and $\phi _{js}$ are, respectively, the
intensity and the phase values corresponding to the analyzed stationary
solution. Then we obtain the linearized Langevin equations describing the
evolution of quantum fluctuations, $\delta \mathbf{a}=\func{col}(\delta
\alpha _{1},\delta \alpha _{1}^{+},\delta \alpha _{2},\delta \alpha
_{2}^{+}) $, which read 
\end{subequations}
\begin{equation}
\frac{d}{dt}\delta \mathbf{a}=\bar{A}\cdot \delta \mathbf{a}+\bar{B}\cdot 
\mathbf{\xi }\left( t\right) ,  \label{LinearizedLangevin}
\end{equation}%
where matrix $\tilde{A}$ corresponds to 
\begin{equation}
\bar{A}_{ij}=\left( \frac{\partial A_{i}}{\partial \ \delta a_{j}}\right) _{%
\mathbf{a}=\mathbf{a}_{s}},
\end{equation}%
and $\bar{B}=B\left( \mathbf{a}=\mathbf{a}_{s}\right) $.

Quantum fluctuations are well characterized by their spectrum \cite%
{Chaturvedi77,Collet} (i.e., by the Fourier transform of the two--time
correlation functions), which are given by the spectral matrix $\mathbf{M}%
\left( \omega \right) $ defined as%
\begin{equation}
\mathbf{M}\left( \omega \right) =\int_{-\infty }^{+\infty }dt\ e^{i\omega
t}\left\langle \delta \mathbf{a}_{i}\left( 0\right) \delta \mathbf{a}%
_{j}\left( t\right) \right\rangle .
\end{equation}%
Chaturvedi et al. \cite{Chaturvedi77} showed that $\mathbf{M}\left( \omega
\right) $ can be directly obtained from Eq.(\ref{LinearizedLangevin}) 
\begin{equation}
\mathbf{M}\left( \omega \right) =\left( \bar{A}+i\omega \mathbb{I}\right)
^{-1}\bar{D}\left( \bar{A}^{T}-i\omega \mathbb{I}\right) ^{-1}
\label{SpectralMatrix}
\end{equation}%
with $\bar{D}=\bar{B}\bar{B}^{T}$ and $\mathbb{I}$ the identity matrix. We
do not give the explicit expression of the matrix elements of $\mathbf{M}%
\left( \omega \right) $ because they are very lengthy.

Now, as we are interested in the quantum fluctuations of the fields outside
the cavity, we must use the input--output theory for calculating the
spectrum of quantum fluctuations of the output fields. As we have used the
generalized-P representation and the cavity is pumped by a coherent field,
the input--output theory \cite{Collet} shows that 
\begin{equation}
\mathbf{M}^{out}\left( \omega \right) =\int_{-\infty }^{+\infty }dt\
e^{i\omega t}\left\langle \delta \mathbf{a}_{i}^{out}\left( 0\right) \delta 
\mathbf{a}_{j}^{out}\left( t\right) \right\rangle =2\gamma \mathbf{M}\left(
\omega \right) .  \label{SpectralMatrixout}
\end{equation}

We write down now expressions for the quadratures' squeezing spectra. We
define the quadratures corresponding to the quantum fluctuations as%
\begin{equation}
\delta X_{j,\beta }^{out}(t)=\delta \alpha _{j}^{out}(t)e^{i\beta }+\delta
\alpha _{j}^{+\ out}(t)e^{-i\beta },  \label{quadrature}
\end{equation}%
where we remind that $j=1,2$ denotes the parallel ($j=1$) or orthogonal ($%
j=2 $) field and $\beta $ is the (arbitrary) quadrature angle.

Now we define the output quadrature squeezing spectra as%
\begin{equation}
:q_{j}^{out}(\beta ,\omega ):\,=\int_{-\infty }^{\infty }\left\langle
:\delta X_{j,\beta }^{out}(t)\delta X_{j,\beta }^{out}(t+\tau
):\right\rangle e^{i\omega \tau }d\tau ,  \label{DefSqueezing}
\end{equation}%
where $:$ $:$ denotes normal and time ordering. By making use of the
input--output formalism and using the spectral matrix (\ref%
{SpectralMatrixout}) the output quadrature squeezing spectra can be written
as 
\begin{subequations}
\begin{eqnarray}
&:&q_{1}^{out}(\beta ,\omega ):\,=2\gamma \left( M_{11}^{out}e^{-i2\beta
}+M_{22}^{out}e^{i2\beta }\right)  \notag \\
&&\ \ \ \ +2\gamma \left( M_{12}^{out}+M_{21}^{out}\right) , \\
&:&q_{2}^{out}(\beta ,\omega ):\,=2\gamma \left( M_{33}^{out}e^{-i2\beta
}+M_{44}^{out}e^{i2\beta }\right)  \notag \\
&&\ \ \ \ +2\gamma \left( M_{34}^{out}+M_{43}^{out}\right) .
\end{eqnarray}%
It will be useful for later purposes to introduce cross--squeezing spectra,
through the generalization of (\ref{DefSqueezing}) 
\end{subequations}
\begin{eqnarray}
&:&q_{jk}^{out}(\beta _{j},\beta _{k},\omega ):\,=  \label{DefCross} \\
&&\ \ \ \ \ \int_{-\infty }^{\infty }\left\langle :\delta X_{j,\beta
_{j}}^{out}(t)\delta X_{k,\beta _{k}}^{out}(t+\tau ):\right\rangle
e^{i\omega \tau }d\tau ,  \notag
\end{eqnarray}%
which are given by 
\begin{subequations}
\begin{eqnarray}
&:&q_{12}^{out}(\beta _{1},\beta _{2},\omega ):\,=2\gamma \left(
M_{13}^{out}e^{-i\sigma _{12}}+M_{24}^{out}e^{i\sigma _{12}}\right)  \notag
\\
&&\ \ \ \ +2\gamma \left( M_{23}^{out}e^{i\delta
_{12}}+M_{14}^{out}e^{-i\delta _{12}}\right) , \\
&:&q_{21}^{out}(\beta _{2},\beta _{1},\omega ):\,=2\gamma \left(
M_{31}^{out}e^{-i\sigma _{21}}+M_{42}^{out}e^{i\sigma _{21}}\right)  \notag
\\
&&\ \ \ \ +2\gamma \left( M_{41}^{out}e^{i\delta
_{21}}+M_{32}^{out}e^{-i\delta _{21}}\right) ,
\end{eqnarray}%
where 
\end{subequations}
\begin{equation}
\sigma _{jk}=\beta _{j}+\beta _{k},\ \ \ \ \delta _{jk}=\beta _{j}-\beta
_{k}.
\end{equation}%
Given the normal ordering, complete absence of fluctuations in a field
quadrature means $:q_{j}^{out}(\beta ,\omega ):\,=-1$. Moreover, by taking
into account the shot noise level outside the cavity it turns out that the
symmetrically ordered squeezing spectra are 
\begin{subequations}
\begin{eqnarray}
q_{j}^{out}(\beta ,\omega ) &=&1\,+\,:q_{j}^{out}(\beta ,\omega ):\,,
\label{shot} \\
q_{21}^{out}(\beta _{2},\beta _{1},\omega ) &=&\,:q_{21}^{out}(\beta
_{2},\beta _{1},\omega ):.
\end{eqnarray}

\section{Analysis of quadrature squeezing at the singlemode solution
bifurcations}

We analyze here quadrature squeezing at the bifurcations affecting the
linearly polarized solution. Following the steps we have outlined, after
some algebra one finds that the quantities of interest at the bifurcations
(where $I_{2s}=0$) can be written as 
\end{subequations}
\begin{subequations}
\begin{align}
& :q_{1}^{out}\left( \beta ,\omega \right) :\,=\frac{4\left\vert
a_{1s}\right\vert ^{2}Q_{1n}}{\left( Q_{1d}-\omega ^{2}\right) ^{2}+4\omega
^{2}},  \label{aux1} \\
& :q_{2}^{out}\left( \beta ,\omega \right) :\,=\frac{6\left\vert
a_{1s}\right\vert ^{2}Q_{2}}{\left( Q_{2d}+2\omega ^{2}\right) ^{2}+16\omega
^{2}},  \label{aux2} \\
& :q_{12}^{out}(\beta _{1},\beta _{2},\omega ):\,=\,:q_{21}^{out}(\beta
_{1},\beta _{2},\omega ):\,=0,
\end{align}%
where 
\end{subequations}
\begin{subequations}
\begin{eqnarray}
Q_{1n} &=&-\left( 3\left\vert a_{1s}\right\vert ^{4}-4\left\vert
a_{1s}\right\vert ^{2}\Delta +\Delta ^{2}-1-\omega ^{2}\right) \sin \psi 
\notag \\
&&+2\left( \Delta -2\left\vert a_{1s}\right\vert ^{2}\right) \cos \psi
+2\left\vert a_{1s}\right\vert ^{2}, \\
Q_{1d} &=&3\left\vert a_{1s}\right\vert ^{4}-4\left\vert a_{1s}\right\vert
^{2}\Delta +\Delta ^{2}+1, \\
Q_{2} &=&\left[ \left\vert a_{1s}\right\vert ^{4}+\Delta \left\vert
a_{1s}\right\vert ^{2}-2\left( \Delta ^{2}-1-\omega ^{2}\right) \right] \sin
\psi  \notag \\
&&+\left( 4\Delta -\left\vert a_{1s}\right\vert ^{2}\right) \cos \psi
+3\left\vert a_{1s}\right\vert ^{2}, \\
Q_{2d} &=&\left\vert a_{1s}\right\vert ^{4}+\Delta \left\vert
a_{1s}\right\vert ^{2}-2\Delta ^{2}-2,
\end{eqnarray}%
and 
\end{subequations}
\begin{equation}
\psi =2\left( \beta -\phi _{1s}\right) ,  \label{psi}
\end{equation}%
is twice the phase difference between the analyzed quadrature and the steady
state phase, and $\left\vert a_{1s}\right\vert ^{2}$ is the normalized
intensity given in Eqs. (\ref{ReNormalizations}) and (\ref{monomode}). In
the above equations the frequency $\omega $ has been normalized to $\gamma $%
, but we do not change the symbol for not complicating unnecessarily the
notation.

Of course a well known general result is that at the bifurcations there is a
field quadrature (of the field directly affected by the bifurcation) that is
perfectly squeezed. A different question is how does the bifurcation
affecting one mode influence the quadrature squeezing of the other
(orthogonal) mode. Let us consider the different bifurcations separately.

\subsection{Polarization bifurcation}

At the polarization bifurcation $I_{1s}=I_{1s,pol}$, Eq. (\ref{ipolarization}%
), and there is a quadrature of field $\alpha _{2}$ (the field that switches
on at this bifurcation) that is perfectly squeezed at frequency $\omega =0$
as expected (see the inset in Fig. 2). It can be shown that the squeezed
quadrature is $\delta X_{2,\beta }^{out}(t)$ with $\beta =\left( \psi
_{pol,opt}/2+\phi _{1s}\right) $ and%
\begin{equation}
\psi _{pol,opt}=-\frac{1}{2}\arccos \left[ \frac{1-\Delta \sqrt{8+9\Delta
^{2}}}{3\left( 1+\Delta ^{2}\right) }\right] ,
\end{equation}%
which tends to zero for large negative $\Delta $ and to $-\pi /2$ for large
positive $\Delta $ (Fig. 2). It is interesting to note that the range of $%
\psi $ values for which squeezing occurs is very narrow.

As for the field $\alpha _{1}$ (the mode that is on) it undergoes large
levels of quadrature noise reduction at this bifurcation, even if the
analyzed bifurcation does not affect the field $\alpha _{1}$ but the field $%
\alpha _{2}$. These levels tend to perfect squeezing for increasing positive 
$\Delta $. In Fig. 3 the squeezing spectrum is shown for two values of the
cavity detuning. The optimum squeezing is reached when $\psi =0$, i.e. for $%
\beta =\phi _{1s}$, i.e., it is an amplitude squeezing, and occurs at the
frequency%
\begin{equation}
\omega _{opt}=\sqrt{5-\frac{7}{2}\Delta \left( -3\Delta +\sqrt{8+9\Delta ^{2}%
}\right) }.
\end{equation}%
In Fig. 4 both the optimum squeezing level and the frequency at which this
squeezing occurs are represented as a function of cavity detuning. At this
stage it is convenient to remember that as $\Delta $ increases, the
polarization and one of the bistability bifurcations approach, see Fig. 1,
what undoubtedly is connected with these large squeezing levels.

\subsection{Bistability bifurcations}

The bistability bifurcation has been studied several times in the past (see,
e.g., \cite{WallsMilburn}) and we do not find it necessary to repeat here
these well known results, specially because we shall not need them when
analyzing polarization squeezing in the following section. It will suffice
to say that perfect squeezing, i.e. $:q_{1}^{out}\left( \beta ,\omega
\right) :\,=-1$, occurs at the bifurcations at $\omega =0$ for a particular
quadrature of the field $\alpha _{1}$. The quadratures that are perfectly
squeezed are 
\begin{subequations}
\begin{eqnarray}
\psi _{up,opt} &=&-\arccos \left[ \frac{2+\Delta \sqrt{\Delta ^{2}-3}}{%
1+\Delta ^{2}}\right] , \\
\psi _{down,opt} &=&-\arccos \left[ \frac{2-\Delta \sqrt{\Delta ^{2}-3}}{%
1+\Delta ^{2}}\right] ,
\end{eqnarray}%
where the subscripts "up" and "down" refer to the tangent (bistability)
bifurcations of the upper and lower branches of the stationary solution,
respectively, which are depicted in Fig. 1. Notice that for $\Delta
\rightarrow \infty $, $\psi _{up,opt}\rightarrow 0$ and $\psi
_{down,opt}\rightarrow \pi $. That is, for $\Delta \rightarrow \infty $ in
the upper (lower) bifurcation there is intensity (phase) squeezing.

Let us now analyze in more detail here the quadrature noise reduction at
these two bifurcation in the other (i.e., the orthogonal) field $\alpha _{2}$%
, which remains off at the pump values at which these bifurcations occur,
see Fig. 1.

\subsubsection{Upper branch}

After substituting $I_{1s}=I_{1s,+}$ in Eq. (\ref{aux2}) one can show that
optimum squeezing occurs for $\psi =\pi $ at a frequency that is very close
to zero for $\Delta $ close to $\sqrt{3}$ (remember that the bistability
bifurcation requires $\Delta \geq \sqrt{3}$) and becomes zero for $\Delta
>1.89$, see Fig. 5. The squeezing level increases with detuning, as can be
seen in Fig. 6, passing from $:q_{2}^{out}\left( \pi ,\omega _{opt}\right)
:\,=-0.75$ at $\Delta =\sqrt{3}$ to $:q_{2}^{out}\left( \pi ,\omega
_{opt}\right) :\,=-0.98$ for $\Delta $ tending to infinity. It is important
to stress for later purposes that optimum squeezing occurs at $\psi =\pi $,
i.e., for $\beta =\phi _{1s}+\pi /2$.

\subsubsection{Lower branch}

After substituting $I_{1s}=I_{1s,-}$ in Eq. (\ref{aux2}) one can show that
optimum squeezing occurs for $\psi =\pi $ at $\omega _{opt}=\sqrt{\frac{1}{18%
}\left[ 7\Delta \left( \Delta +\sqrt{\Delta ^{2}-3}\right) -15\right] }$, a
frequency that increases linearly with $\Delta $ from its minimum value $%
\omega =1/\sqrt{3}$ at $\Delta =\sqrt{3}$. Fig. 7 shows the squeezing
spectrum for two values of the cavity detuning, and Fig. 8 shows the optimum
squeezing as well as the frequency at which it occurs. The squeezing level
is not perfect (the optimum is $:q_{2}^{out}\left( \pi ,\omega _{opt}=1/%
\sqrt{3}\right) :\,=-0.75$ for $\Delta =\sqrt{3}$), and degrades with
increasing detuning. As in the previous case, we stress that optimum
squeezing occurs at $\psi =\pi $, i.e., for $\beta =\phi _{1s}+\pi /2$.

\section{Quantum Stokes parameters}

The polarization state of light is classically described by using the Stokes
parameters \cite{BornWolf}. The Hermitean Stokes operators are defined
directly from the analogy with the classical parameters \cite{Korolkova02} 
\end{subequations}
\begin{subequations}
\label{StokesParameters}
\begin{eqnarray}
\hat{S}_{0} &=&\hat{a}_{1}^{\dagger }\hat{a}_{1}+\hat{a}_{2}^{\dagger }\hat{a%
}_{2},\ \ \ \hat{S}_{1}=\hat{a}_{1}^{\dagger }\hat{a}_{1}-\hat{a}%
_{2}^{\dagger }\hat{a}_{2}, \\
\hat{S}_{2} &=&\hat{a}_{1}^{\dagger }\hat{a}_{2}+\hat{a}_{2}^{\dagger }\hat{a%
}_{1},\ \ \ \hat{S}_{3}=i\left( \hat{a}_{2}^{\dagger }\hat{a}_{1}-\hat{a}%
_{1}^{\dagger }\hat{a}_{2}\right) .
\end{eqnarray}%
Operator $\hat{S}_{0}$ refers to the beam intensity, whilst $\hat{S}_{1},%
\hat{S}_{2}$ and $\hat{S}_{3}$ describe the polarization state. The
parameter $\hat{S}_{0}$ commutes with all the others 
\end{subequations}
\begin{equation}
\left[ \hat{S}_{0},\hat{S}_{j}\right] =0~,~j=1,2,3  \label{CommutatorS0}
\end{equation}%
whereas the remaining parameters satisfy the commutation relation of the
SU(2) Lie algebra%
\begin{equation}
\left[ \hat{S}_{k},\hat{S}_{l}\right] =2i\varepsilon _{klm}\hat{S}_{m}.
\label{CommutatorSU}
\end{equation}%
Therefore, the simultaneous measurements of the Stokes parameters are
impossible in general. The non--zero commutator in (\ref{CommutatorSU})
implies a restriction to the variances of the Stokes operators, in the form
of the uncertainty relations%
\begin{equation}
V_{k}V_{l}\geq \left\vert \left\langle \hat{S}_{m}\right\rangle \right\vert
^{2},\ l~\neq m\neq k,\ \ l\neq k,  \label{Uncertainty}
\end{equation}%
where 
\begin{equation}
V_{k}=\left\langle \hat{S}_{k}^{2}\right\rangle -\left\langle \hat{S}%
_{k}\right\rangle ^{2},
\end{equation}%
is the variance of the quantum Stokes parameter $\hat{S}_{k}$. We find it
convenient to normalize the variances to the mean intensity and thus we
shall use 
\begin{equation}
\tilde{V}_{k}=\frac{\left\langle \hat{S}_{k}^{2}\right\rangle -\left\langle 
\hat{S}_{k}\right\rangle ^{2}}{\left\langle \hat{S}_{0}\right\rangle },
\end{equation}%
and then the inequality (\ref{Uncertainty}) reads 
\begin{equation}
\tilde{V}_{k}\tilde{V}_{l}\geq \left\vert \frac{\left\langle \hat{S}%
_{m}\right\rangle }{\left\langle \hat{S}_{0}\right\rangle }\right\vert
^{2},\ l~\neq m\neq k.  \label{Uncertaintynorm}
\end{equation}%
We stress that the uncertainty relations (\ref{Uncertainty}) depend on the
expected value of the operators, which makes non-trivial the definition of
polarization squeezed states \cite{Heersink03,Luis06}.

In terms of field quadratures, a squeezed state is that which (a) is a
minimum uncertainty state (MUS) and (b) has fluctuations in a particular
field quadrature below those corresponding to the vacuum state (or a
coherent state as it is synonymous). Obviously the reduction of fluctuations
in one field quadrature occurs at the expense of increasing the orthogonal
field quadrature fluctuations. In the case of polarization squeezing this is
no more the case because of the already commented appearance of the expected
value in inequality (\ref{Uncertainty}). Let us see this is some more
detail. A coherent state verifies \cite{Heersink03} $\tilde{V}_{k}=1$ for
all the Stokes parameters but a coherent state is not, in general, a MUS in
terms of polarization (this depends on its polarization state): In effect,
consider a coherent state such that for some $m$ it verifies $\left\langle 
\hat{S}_{m}\right\rangle <\left\langle \hat{S}_{0}\right\rangle $, then
obviously $\tilde{V}_{k}\tilde{V}_{l}>\left\vert \left\langle \hat{S}%
_{m}\right\rangle /\left\langle \hat{S}_{0}\right\rangle \right\vert ^{2}$
for this coherent state as $\tilde{V}_{k}\tilde{V}_{l}=1$. It is also
obvious that there can be states with some $\tilde{V}_{k}<1$ which are not
polarization squeezed states: Think of a state with $\left\langle \hat{S}%
_{m}\right\rangle =0$ for some $m$ and $\tilde{V}_{k}<1$. It is clear that
is not experiencing any transfer of fluctuations from one Stokes parameter
to another one, any rearrangement of quantum fluctuations. Then the quantum
vacuum is not a reference for defining the squeezed polarization state.

These are the reasons why polarization squeezing is defined in a different
way:\ A field state is said to be squeezed if \cite{Heersink03} 
\begin{equation}
\tilde{V}_{l}<\left\vert \frac{\left\langle \hat{S}_{m}\right\rangle }{%
\left\langle \hat{S}_{0}\right\rangle }\right\vert <\tilde{V}_{k}~;\ \ \
l~\neq m\neq k\neq 0.  \label{sq}
\end{equation}%
for some $l$, i.e., a polarization squeezed state is a state for which the
variance of one of the Stokes parameters lies not only below the coherent
limit, but also below the corresponding minimum uncertainty state limit. In
a state verifying such a condition there is a rearrangement of quantum
fluctuations between the different Stokes parameters and is thus a proper
squeezed state.

We shall adopt the above criterion (\ref{sq}) but we must mention that there
has been some debate concerning the most appropriate criterion for
polarization squeezing and we refer the reader to \cite%
{Korolkova02,Heersink03,Luis06} and references therein.

\section{Stokes parameters' fluctuation spectra}

As stated, the polarization squeezing properties of light are provided by
the knowledge of the variances of the quantum Stokes operators. As we
commented in the previous section, polarization and quadrature squeezing are
not equivalent, and a quadrature squeezed state does not imply necessarily a
polarized squeezed state. Nevertheless, we find it much convenient to write
down the spectra of the variances of the quantum Stokes parameters in terms
of the quadrature squeezing spectra (\ref{DefSqueezing}) and (\ref{DefCross}%
), showing up the relation between them. We see below that the cross-mode
spectra (\ref{DefCross}) differentiates the polarization squeezing from a
simple combination of quadrature spectra.

After lengthy but straightforward algebra one obtains that the variances of
the quantum Stokes operators\ can be written as 
\begin{subequations}
\begin{eqnarray}
V_{0}(\omega ) &=&I_{1s}q_{1}^{out}(\phi _{1s},\omega
)+I_{2s}q_{2}^{out}(\phi _{2s},\omega )  \notag \\
&&+\sqrt{I_{1s}I_{2s}}q_{12}^{out}(\phi _{1s},\phi _{2s},\omega )  \notag \\
&&+\sqrt{I_{1s}I_{2s}}q_{21}^{out}(\phi _{2s},\phi _{1s},\omega ),
\end{eqnarray}%
\begin{eqnarray}
V_{1}(\omega ) &=&I_{1s}q_{1}^{out}(\phi _{1s},\omega
)+I_{2s}q_{2}^{out}(\phi _{2s},\omega )  \notag \\
&&-\sqrt{I_{1s}I_{2s}}q_{12}^{out}(\phi _{1s},\phi _{2s},\omega )  \notag \\
&&-\sqrt{I_{1s}I_{2s}}q_{21}^{out}(\phi _{2s},\phi _{1s},\omega ),
\end{eqnarray}

\begin{eqnarray}
V_{2}(\omega ) &=&I_{2s}q_{1}^{out}\left( \phi _{2s},\omega \right)
+I_{1s}q_{2}^{out}\left( \phi _{1s},\omega \right)  \notag \\
&&+\sqrt{I_{1s}I_{2s}}q_{12}^{out}\left( \phi _{2s},\phi _{1s},\omega \right)
\notag \\
&&+\sqrt{I_{1s}I_{2s}}q_{21}^{out}\left( \phi _{1s},\phi _{2s},\omega
\right) ,  \label{v2}
\end{eqnarray}%
\begin{eqnarray}
V_{3}(\omega ) &=&I_{2s}q_{1}^{out}\left( \phi _{2s}+\tfrac{\pi }{2},\omega
\right) +I_{1s}q_{2}^{out}\left( \phi _{1s}+\tfrac{\pi }{2},\omega \right) ,
\notag \\
&&-\sqrt{I_{1s}I_{2s}}q_{12}^{out}\left( \phi _{2s}+\tfrac{\pi }{2},\phi
_{1s}+\tfrac{\pi }{2},\omega \right)  \notag \\
&&-\sqrt{I_{1s}I_{2s}}q_{21}^{out}\left( \phi _{1s}+\tfrac{\pi }{2},\phi
_{2s}+\tfrac{\pi }{2},\omega \right) ,
\end{eqnarray}%
where $I_{is}$ and $\phi _{is}$ correspond to the steady state values given
by Eqs. (\ref{monomode}) and Eqs. (\ref{Bimode}) for the singlemode and
bimode solutions, respectively. As for the quadrature squeezing spectra
elements appearing in the above equations, they have been written
accordingly to the definitions in Eqs. (\ref{DefSqueezing}) and (\ref%
{DefCross}).

Next we analyze the behavior of $\tilde{V}_{k}(\omega )=V_{k}(\omega
)/\left\langle \hat{S}_{0}\right\rangle $, with $\left\langle \hat{S}%
_{0}\right\rangle =I_{1s}+I_{2s}$. We do this first at the bifurcations we
have analyzed in Section 3, and then we analyze $\tilde{V}_{k}(\omega )$ as
a function of pump for a particular value of the detuning.

\subsection{Polarization squeezing at the bifurcations}

Eqs. (\ref{v}) simplify a lot at the bifurcation points (\ref%
{bistabilityLoop}) and (\ref{BimodeSwitch}) as at these particular pump
values the steady state solution is the singlemode one, Eq. (\ref{monomode}%
). Then, setting $I_{2s}=0$ one immediately obtains 
\end{subequations}
\begin{subequations}
\label{VStokesmono}
\begin{eqnarray}
\tilde{V}_{0}(\omega ) &=&\tilde{V}_{1}(\omega )=1+\,:q_{1}(\phi
_{1s},\omega ):, \\
\tilde{V}_{2}(\omega ) &=&1+\,:q_{2}\left( \phi _{1s},\omega \right) :, \\
\tilde{V}_{3}(\omega ) &=&1+\,:q_{2}\left( \phi _{1s}+\tfrac{\pi }{2},\omega
\right) :,
\end{eqnarray}%
where we have taken into account Eq. (\ref{shot}).

The above expressions show that in the singlemode solution the variances of
the two first Stokes parameters correspond to the quadrature fluctuations of
the emitting (parallel) mode $\alpha _{1}$ at the particular quadrature
angle $\beta =\phi _{1s}$ (i.e., the intensity fluctuations as it must be)
which, in general, is not the quadrature that exhibits less fluctuations.
Contrarily, the two last Stokes parameters correspond to the quadrature
fluctuations of the non--emitting (orthogonal) mode $\alpha _{2}$ at the
particular quadrature angles $\beta =\phi _{1s}$ and $\beta =\phi _{1s}+%
\frac{\pi }{2}$.

In this case, i.e., at the bifurcations, we can give the explicit
expressions of the $\tilde{V}_{i}(\omega )$ that read 
\end{subequations}
\begin{subequations}
\begin{eqnarray}
\tilde{V}_{0}(\omega ) &=&\tilde{V}_{1}(\omega )=1+\frac{4\left\vert
a_{1s}\right\vert ^{2}\left( \Delta -\left\vert a_{1s}\right\vert
^{2}\right) }{C_{1}}, \\
\tilde{V}_{2}(\omega ) &=&1+\frac{6\left\vert a_{1s}\right\vert ^{2}\left(
\left\vert a_{1s}\right\vert ^{2}+2\Delta \right) }{C_{2}}, \\
\tilde{V}_{3}(\omega ) &=&1+\frac{2\left( \left\vert a_{1s}\right\vert
^{2}-\Delta \right) }{\left( \left\vert a_{1s}\right\vert ^{2}+2\Delta
\right) }\tilde{V}_{2}(\omega ),
\end{eqnarray}%
where 
\end{subequations}
\begin{subequations}
\begin{eqnarray}
C_{1} &=&\left( \omega ^{2}+1-3\left\vert a_{1s}\right\vert ^{4}+4\left\vert
a_{1s}\right\vert ^{2}\Delta -\Delta ^{2}\right) ^{2} \\
&&+4\left( 3\left\vert a_{1s}\right\vert ^{4}-4\left\vert a_{1s}\right\vert
^{2}\Delta +\Delta ^{2}\right) ,  \notag \\
C_{2} &=&\left( 2\omega ^{2}+2+\left\vert a_{1s}\right\vert ^{4}+\left\vert
a_{1s}\right\vert ^{2}\Delta -2\Delta ^{2}\right) ^{2} \\
&&-8\left( \left\vert a_{1s}\right\vert ^{2}-\Delta \right) \left(
\left\vert a_{1s}\right\vert ^{2}+2\Delta \right) ,  \notag
\end{eqnarray}%
with $\Delta =\theta /\gamma \eta $ and $\left\vert a_{1s}\right\vert
^{2}=gI_{1s}/\gamma $, as introduced in Eq. (\ref{a}), and where $\omega $
is normalized to $\gamma $. Now the above expressions must be particularized
for the different possible bifurcations affecting the singlemode solution.

We must now take into account that as $I_{2s}=0$ at the bifurcations, at
these points the mean values of the Stokes parameters are $\left\langle \hat{%
S}_{0}\right\rangle =\left\langle \hat{S}_{1}\right\rangle =I_{1s}$ and $%
\left\langle \hat{S}_{2}\right\rangle =\left\langle \hat{S}_{3}\right\rangle
=0$, and then Eq. (\ref{sq}) implies that the only Stokes parameters that
can be squeezed are $\tilde{V}_{2}$ and $\tilde{V}_{3}$, and for that they
must verify 
\end{subequations}
\begin{equation}
\tilde{V}_{i}<\left\vert \frac{\left\langle \hat{S}_{1}\right\rangle }{%
\left\langle \hat{S}_{0}\right\rangle }\right\vert <\tilde{V}_{j},\ \ \
i,j=2,3,\ \ i\neq j.
\end{equation}

\subsubsection{Polarization bifurcation}

In this case $I_{1s}=I_{1s,pol}$, Eq. (\ref{ipolarization}) and it is not
difficult to show that neither $\tilde{V}_{2}$ nor $\tilde{V}_{3}$ are
squeezed. We saw in Subsection 5.1 that optimum quadrature squeezing occurs
at this bifurcation for $\beta =\left( \psi _{pol,opt}/2+\phi _{1s}\right) $
with $\psi _{pol,opt}$ tending to $0$ for $\Delta \rightarrow -\infty $ and
to $-\pi /2$ for $\Delta \rightarrow \infty $. It is immediately seen from
this that only for $\Delta \rightarrow -\infty $ it could happen that $%
\tilde{V}_{2}$ be squeezed, see Eqs. (\ref{VStokesmono}) and Eq. (\ref{psi}%
), but it turns out that this never occurs for finite values of the detuning
because, as commented in Subsection 5.1, the squeezing level is extremely
sensitive to the $\psi $ value. Then there is not any polarization squeezing
at this bifurcation.

\subsubsection{Bistability bifurcation}

In this case $I_{1s}=I_{1s,\pm }$, Eq. (\ref{ibistability}). The fact the
optimum quadrature squeezing for the field $\alpha _{2}$ occurs for $\beta
=\phi _{1s}+\pi /2$ , see Subsection 5.2, makes that $\tilde{V}_{3}$ be
squeezed, see Eqs. (\ref{VStokesmono}). Then in this particular case
quadrature squeezing implies polarization squeezing. In Fig. 9 we represent
the optimum (minimum)\ value of $\tilde{V}_{3}$ at the two bistability
bifurcations, which occurs at the frequencies given in Subsection 5.2.

\subsection{Polarization squeezing as a function of pump}

In this subsection we study the dependence of fluctuations on pump intensity
for a particular value of the detuning. We have chosen $\Delta =1$ as for
this value the only bifurcation suffered by the singlemode solution is the
polarization bifurcation (remember that $\Delta >\sqrt{3}$ for bistability
exists) and thus there is not polarization squeezing at the bifurcation
(Section VII.A.1). Furthermore, from \cite{Sanchez-Morcillo00} we know that
for $\Delta =1$ there is not any secondary bifurcation affecting the bimode
solution.

We concentrate on the fluctuations of the Stokes parameters of the bimode
solution. In Fig. 10 the steady state intensity as well as the mean values
of the Stokes parameters are represented as a function of pump, and in Fig.
11 the minimum value of the variances of the Stokes parameters are
represented together with the mean values of the Stokes parameters. We do
not find it necessary to indicate at which frequency the minimum value of
the variances is reached. Notice that polarization squeezing occurs when the
represented variances fall below the mean value of the appropriate Stokes
parameter, which is the one that is shown in each figure, see Eq. (\ref{sq}).

We see that in spite of the fact that no polarization squeezing exists at
the bifurcation, as commented in the previous subsection, as pump is
increased parameters $S_{1}$ and $S_{2}$ are squeezed within some pump value
ranges (we note that $V_{3}$ increases its value for pump values larger than
those represented in the figure). The observed polarization squeezing must
be attributed in this case to the effect of the cross--mode spectra in Eqs. (%
\ref{v}). We would like to note that in spite of the fact that there is
always some parameter for which $\tilde{V}_{i}<1$, see Fig. 11. Although
this is not polarization squeezing \cite{Heersink03}, it is a fact that the
output field exhibits less fluctuations than a coherent state and this could
also be useful as it is a form of intensity squeezing.

These results are shown just to illustrate that the vectorial Kerr cavity
exhibits polarization squeezing not only at the bifurcations and we have
checked that this happens for other parameter values. As the behavior of the
bimode solution can be quite involved we do not find it necessary to go in
our analysis beyond this point.

\section{Conclusions}

In this article we have studied quadrature and polarization squeezing in a
vectorial Kerr cavity model. After deriving Langevin equations for the
quantum fluctuations from a suitable Fokker--Planck equation, namely that of
the generalized P--distribution, we have derived expressions for the quantum
fluctuations of the output field quadratures as a function of which we have
expressed the spectrum of fluctuations of the output field Stokes
parameters. These last expressions are particularly useful for understanding
the connection between quadrature and polarization squeezing and facilitate
the study of the latter.

Through the analysis of the quadrature and polarization fluctuations
spectra, we have analyzed the conditions under which squeezing occur. We
have concentrated mainly on the squeezing occurring at the bifurcations
affecting the singlemode (linearly polarized) solution, but have considered
also polarization squeezing in the bimode (elliptically polarized) solution
in a particular case. In particular we have demonstrated that polarization
squeezing occurs in this system, being specially large in the upper tangent
(bistability) bifurcation of the singlemode solution.

Although the results we have shown have been derived for particular values
of the Maker--Terhune coefficients (namely corresponding to the case of
liquids), we have analyzed also other cases and have convinced ourselves
that our results are qualitatively general: Except for the case $\mathcal{B}%
=0$ (that describes isotropic media in which the Kerr nonlinearity is solely
due to electrostiction) in which the field $\alpha _{2}$ is a coherent
vacuum irrespective of the pump value (the polarization instability does not
exists in this case), for any $\mathcal{B}\neq 0$ there exists polarization
squeezing at the tangent bifurcations (of course the amount of squeezing
does depend on the particular $\mathcal{B}$ value).

From these results we conclude that the vectorial Kerr cavity model
(including both cross--phase modulation and four--wave mixing of the two
orthogonally polarized intracavity modes) constitutes a particularly
interesting generalization of the standard dispersive bistability model from
the point of view of quantum fluctuations, as it predicts the existence of
polarization squeezing, a phenomenon that, obviously, cannot be described
with the standard model, and that exists whenever four--wave mixing occurs
(i.e., whenever $\mathcal{B}\neq 0$). It is particularly interesting that
polarization squeezing occurs in the singlemode solution, being optimum at
the tangent (bistability) bifurcations, i.e. without the need of reaching
the polarization bifurcation.

This work has been supported by the Spanish Ministerio de Educaci\'{o}n y
Ciencia (M.E.C.) and the European Union FEDER through Projects
FIS2005-07931-C03-01 and -02. I.P.--A. also acknowledges finantial support
from the M.E:C. \textit{Juan de la Cierva }programme.

\bigskip

{\LARGE Figure Captions}

\textbf{Fig. 1.} (a)\ Representation on the $\left\langle E^{2},\Delta
\right\rangle $ plane of the two tangent bifurcations (full lines)\ and the
polarization bifurcation (dashed line). The lower (down) and upper (up)
tangent bifurcations receive these names because they occur, respectively,
in the lower and upper branches of the hysteresis cycle. In (b) the output
intensities are shown as a function of pump $E^{2}$ for $\Delta =2$. The
dashed lines in (b) indicate unstable solutions. Another example of
stationary solutions is shown, later, in Fig. 10(a).

\textbf{Fig. 2.} The optimally squeezed quadrature of the field $\alpha _{2}$
(orthogonally polarized with respect to the pump) at the polarization
bifurcation is $\beta =\left( \psi _{pol,opt}/2+\phi _{1s}\right) $. $\psi
_{pol,opt}$ is represented as a function of detuning. In the inset, the
squeezing spectrum of this quadrature is represented for $\Delta =0.$

\textbf{Fig. 3.} Squeezing spectrum of the field $\alpha _{1}$ (the mode
with parallel polarization with respect to that of the pump) of the
optimally squeezed quadrature at the polarization bifurcation for the two
values of the cavity detuning indicated in the figure.

\textbf{Fig. 4.} Optimum squeezing level (full line, left vertical axis) and
the frequency at which this squeezing occurs (dashed line, right vertical
axis) for the field $\alpha _{1}$ at the polarization bifurcation.

\textbf{Fig. 5.} Squeezing spectrum of the field $\alpha _{2}$ (the mode
with orthogonal polarization with respect to that of the pump) of the
optimally squeezed quadrature at the upper tangent bifurcation suffered by
the field $\alpha _{1}$, for the two values of the cavity detuning indicated
in the figure.

\textbf{Fig. 6.} Optimum squeezing level for the field $\alpha _{2}$ at the
upper tangent bifurcation.

\textbf{Fig. 7.} As Fig. 5 but for the lower tangent bifurcation suffered by
the field $\alpha _{1}$, for the two values of the cavity detuning indicated
in the figure.

\textbf{Fig. 8.} Optimum squeezing level (full line, left vertical axis) and
the frequency at which this squeezing occurs (dashed line, right vertical
axis) for the field $\alpha _{2}$ at the lower tangent bifurcation.

\textbf{Fig. 9.} Normalized variance of the $S_{3}$ Stokes parameter at the
upper and lower tangent bifurcations as a function of cavity detuning. The
optimum squeezed value is shown.

\textbf{Fig. 10.} (a)\ Intensity values of the two intracavity modes and (b)
normalized mean value of the Stokes parameters for $\Delta =1$. The
polarization instability is marked with an arrow in the horizontal axis. The
full and dashed lines in (b) correspond to the two possible values of the
Stokes parameters due to the two phase bistability of Eqs. (\ref{clas}) for
the $\alpha _{2}$ field.

\textbf{Fig. 11.} Minimum value of the normalized Stokes parameters'
variances (full lines) and mean values of the Stokes parameters (dashed
lines) as a function of pump intensity for $\Delta =1$. Parameter $S_{2}$ is
squeezed within a small pump parameter domain values around $E^{2}=3$, as
can be seen in (a) and (c), while $S_{1}$ is squeezed for $E^{2}>5$, as can
be seen (c). For larger pump intensity values, polarization squeezing
dissapears.

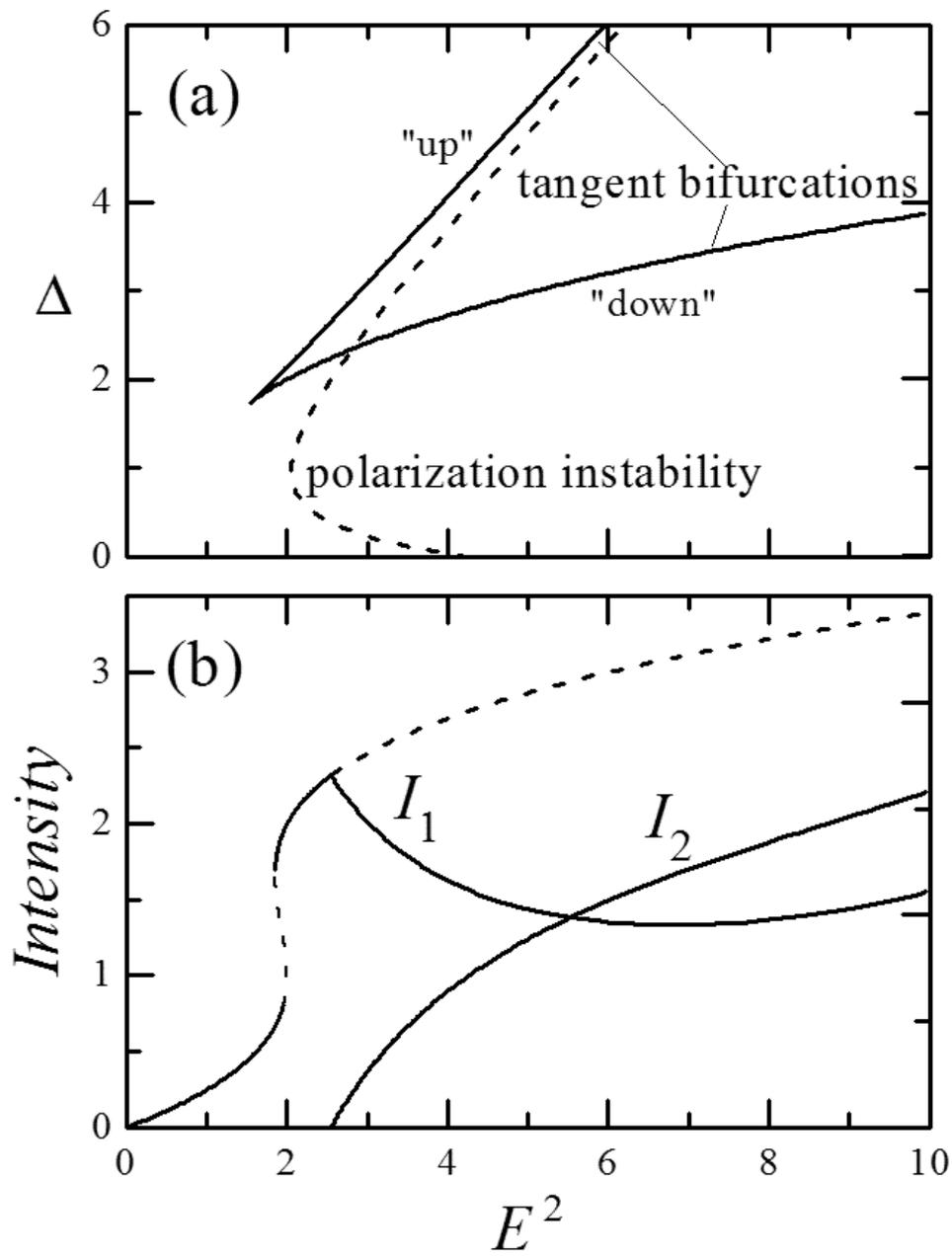
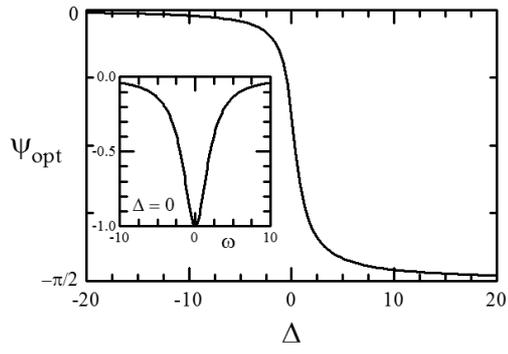
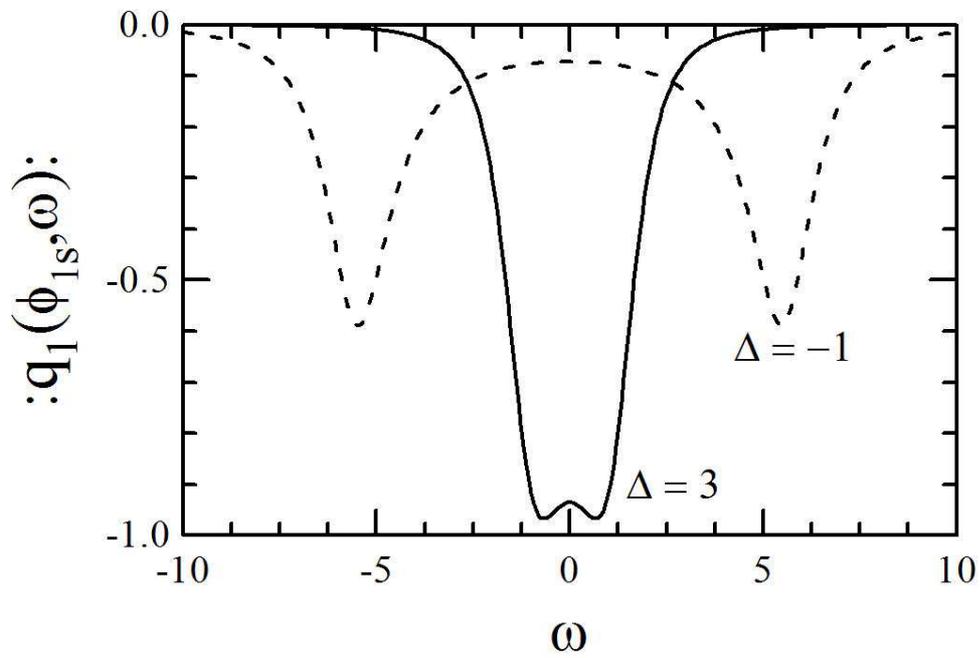
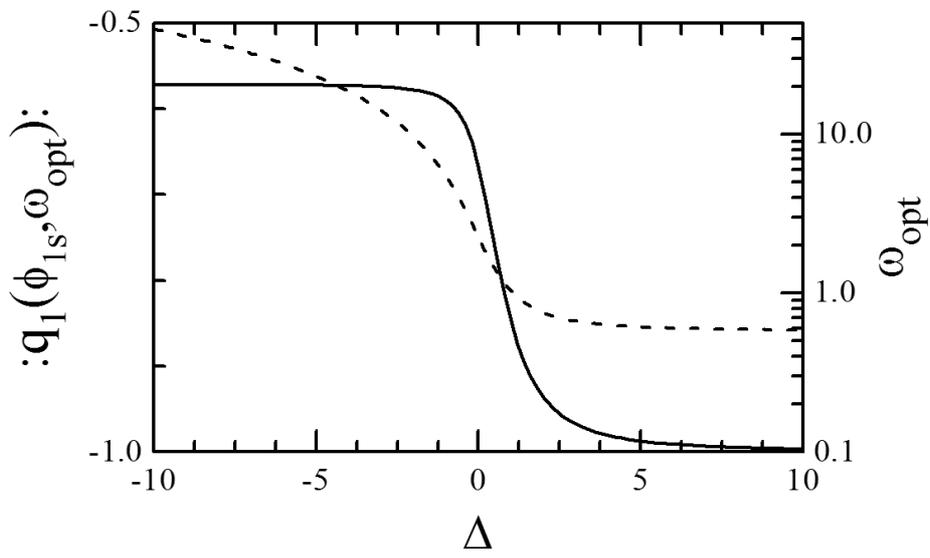
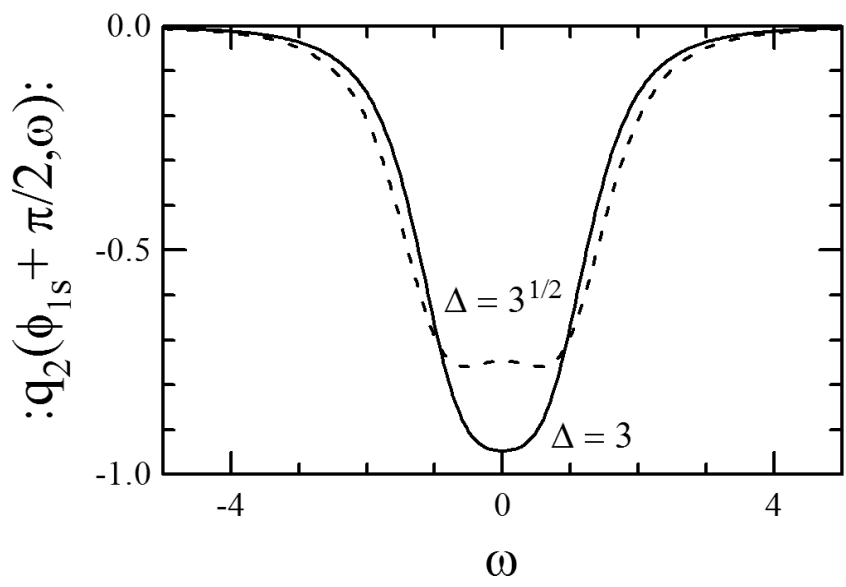
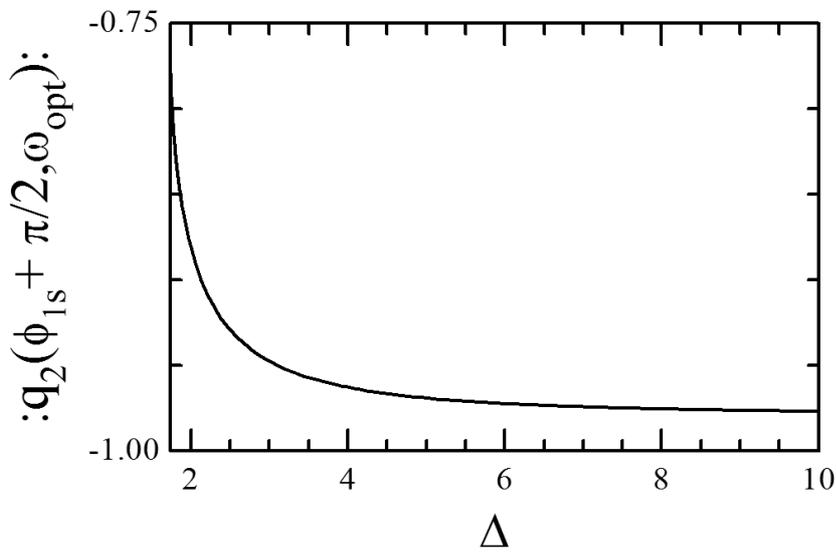
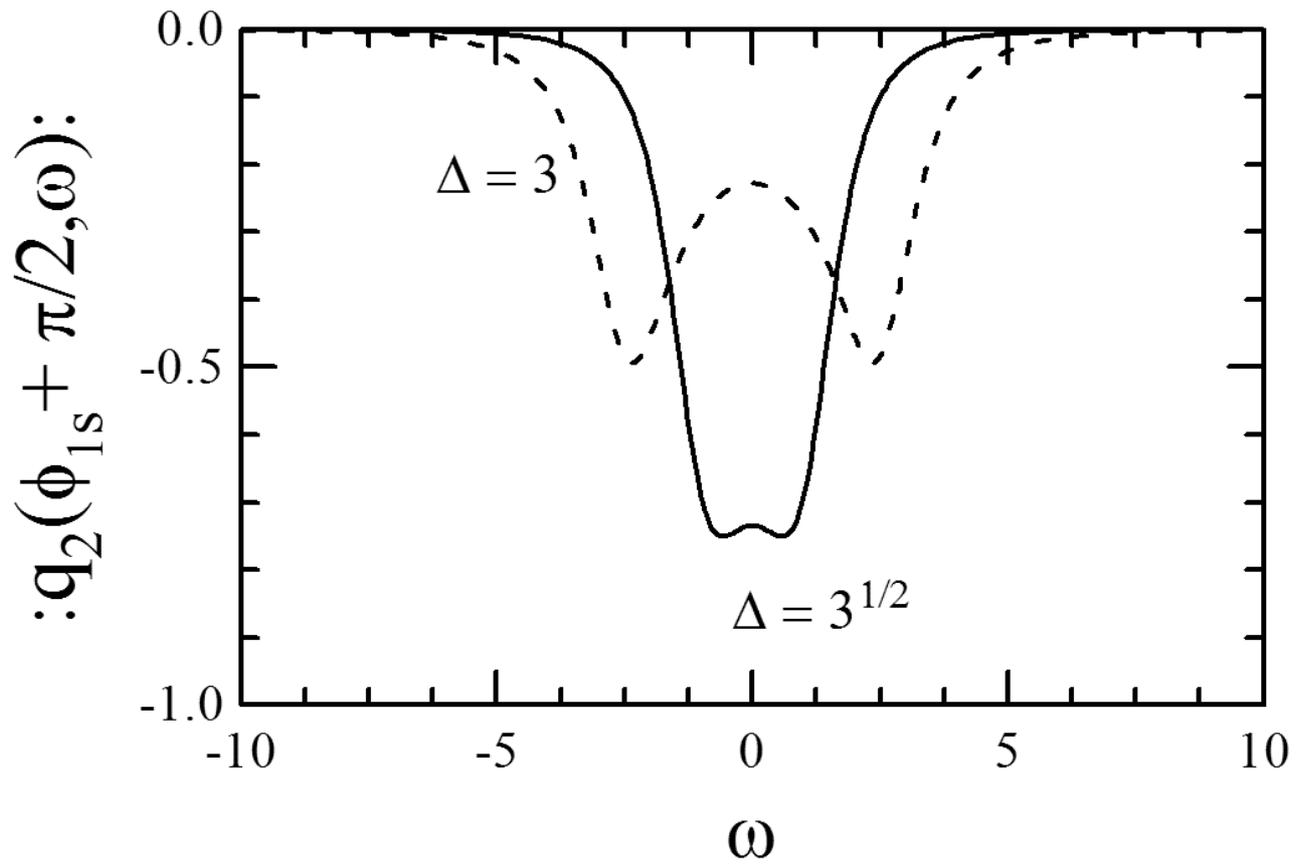
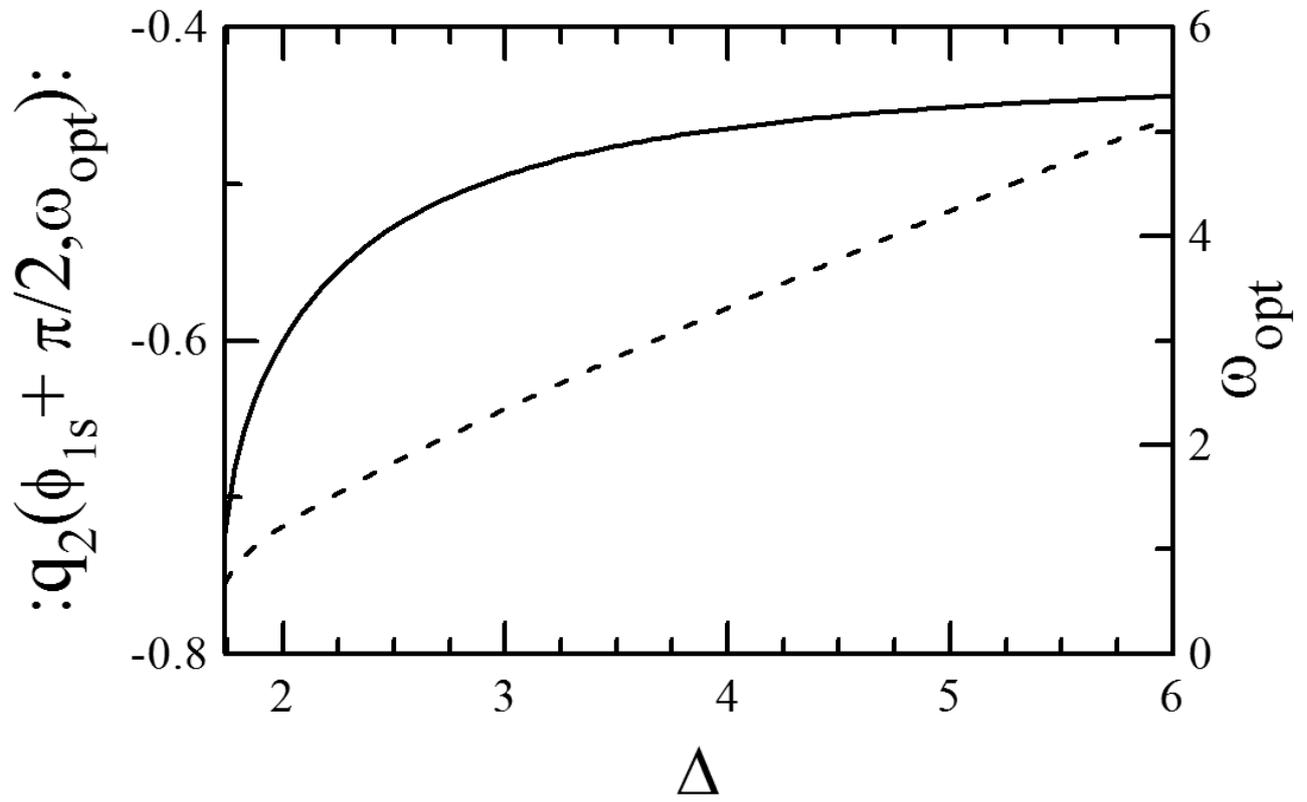
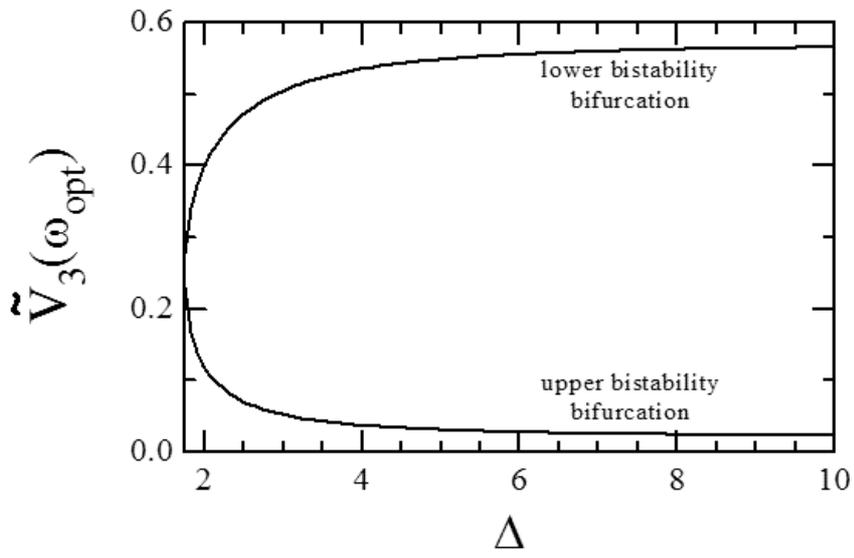
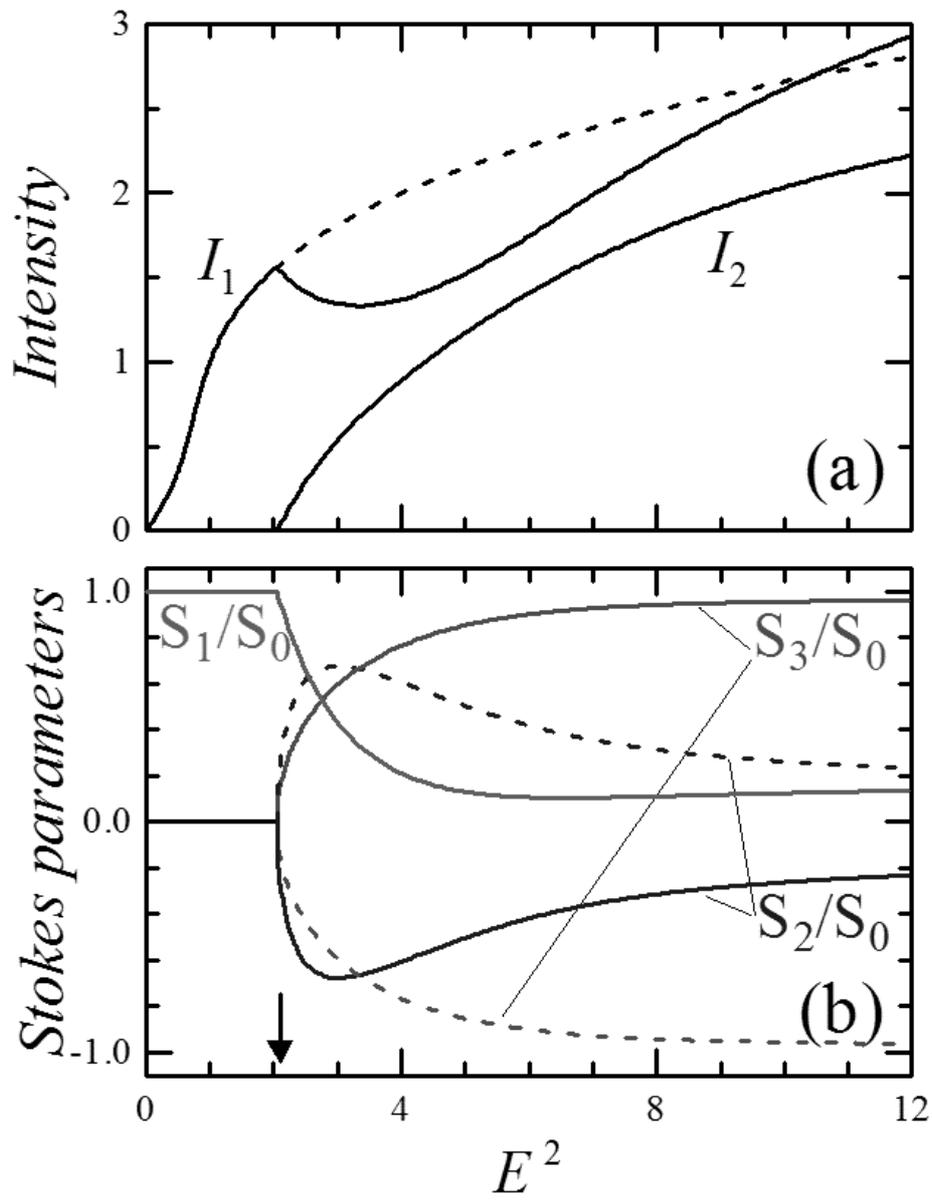
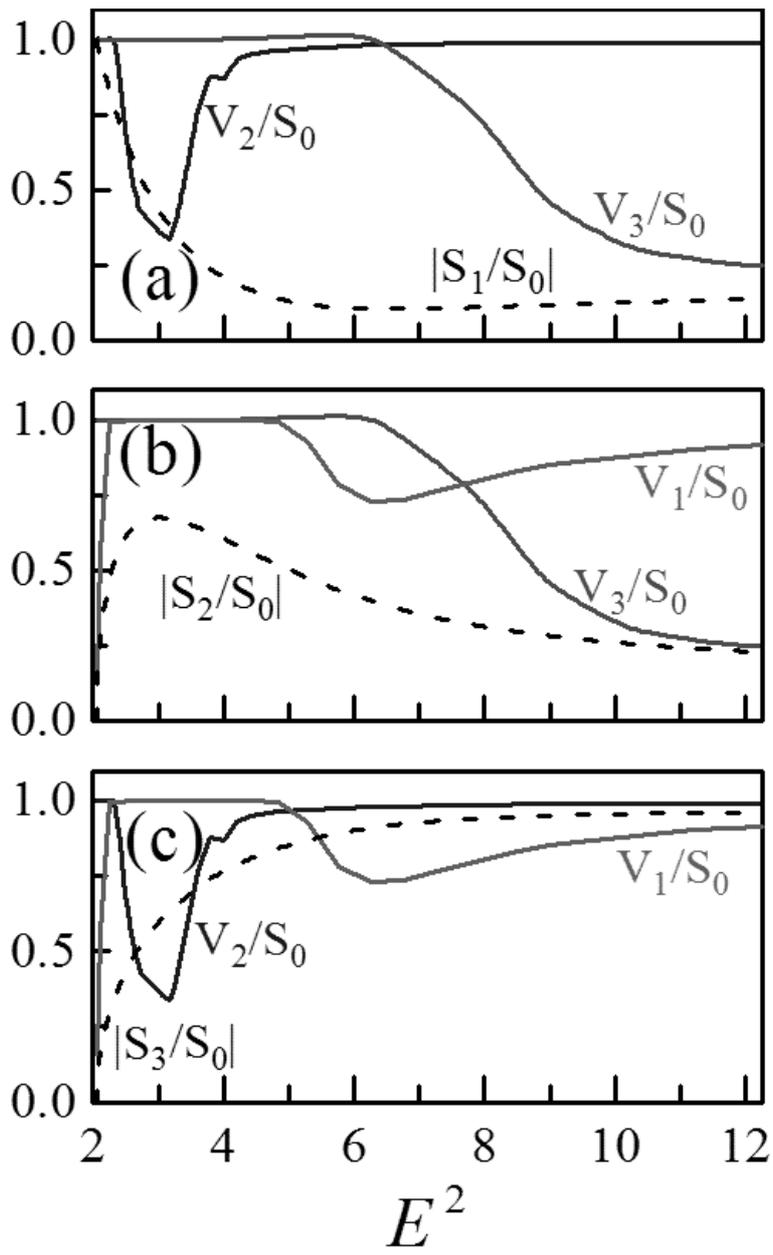
\end{document}